\documentclass[preprint,aps,draft]{revtex4}
\usepackage{graphicx}
\usepackage{amsfonts}
\usepackage{amssymb}

\begin{document}

\preprint{}

\title{Modified dispersion relations and (A)dS Schwarzschild Black holes}

\author{Xin Han}
 \email{bifrostx@gmail.com}
\author{Huarun Li}
 \email{alloys91@sohu.com}
\author{Yi Ling}%
 \email{yling@ncu.edu.cn}
\affiliation{ Center for Relativistic Astrophysics and High Energy
Physics, Department of Physics, Nanchang University, Nanchang
330031, China
}%

\begin{abstract}
In this paper we investigate the impact of modified dispersion
relations (MDR) on (Anti)de Sitter-Schwarzschild black holes.  In
this context we find the temperature of black holes can be derived
with important corrections. In particular given a specific MDR the
temperature has a maximal value such that it can prevent black holes
from total evaporation. The entropy of the (A)dS black holes is also
obtained with a logarithmic correction.
\end{abstract}

\pacs{} \keywords{} \maketitle

\section{Introduction}
\baselineskip=20pt

It is well known that Planck mass $M_p$ or Planck length $l_p$ plays
an important role in quantum gravity. General belief is that the
Planck length may be the minimal observable length
\cite{Gross:1987ar, Maggiore:1993rv}. So it is natural to take the
Planck length as a universal constant. But this seems leading to a
puzzle, saying that the Lorentz symmetry at Planck scale is not
preserved since the length is obviously not an invariant under
linear Lorentz boost. One of approaches to solve this paradox is
non-linear special relativity or doubly special relativity
(DSR)\cite{Magueijo:2001cr}, which may preserve the relativity
principle and at the same time treat Planck energy as an invariant.
In this framework, Lorentz symmetry is deformed such that the usual
energy-momentum relation or dispersion relation may be modified at
Planck scale. As pointed out in \cite{Magueijo:2002am}, a general
modified dispersion relation may be written as
\begin{equation}
E^2f^2_1(E;\eta)-p^2f^2_2(E;\eta)=m^2_0, \label{mdr1}
\end{equation}
where $f_1$ and $f_2$ are two functions of energy from which a
specific formulation of boost generator can be defined. $\eta$ is a
dimensionless parameter(We set $\hbar=c=1$ through the whole paper)
characterizing the strength of the correction. It is also
interesting to see that this formula can be incorporated into a
general relativity framework, see ref. \cite{Magueijo:2002xx}.
Besides this, modified dispersion relations and its implications
have been greatly investigated in recent years, references can be
found in \cite{Colladay98fq, Coleman98ti, Amelino00zs, Jacobson01tu,
Myers03fd, Jacobson03bn, Moffat92ud, Albrecht98ir,
MersiniHoughton:2001su}. In particular, it has been known that MDR
may change the thermodynamical properties of black holes greatly so
as to provide novel mechanism for understanding the late fate of the
black hole evaporation \cite{AmelinoCamelia:2005ik,Ling:2005bq}. In
particular, in \cite{Ling:2005bq} one of our authors with other
collaborators studied the impact of MDR on the thermodynamics of
Schwarzschild black holes. It has been found that due to the
modification of dispersion relations the ordinary picture of Hawking
radiation changes greatly when the mass of black holes approaches to
the Planck scale. First of all, both the temperature and the entropy
of black holes receive important corrections such that the
temperature is bounded with a finite value rather than divergent.
Secondly, such corrections may prevent the black hole from total
evaporation since the heat capacity vanishes as the temperature
reaches the maximal value. Such remnants of black holes may be
viewed as a candidate for dark matter.

In this paper, we intend to extend above analysis to (A)dS
Schwarzschild black holes. We firstly review that the usual Hawking
temperature of (A)dS Schwarzschild black holes can be heuristically
derived by employing the standard dispersion relation as well as
extended uncertainty principle (EUP) to radiation particles in the
vicinity of horizon. Then we investigate how a general form of MDR
may affect the temperature as well as the entropy of black holes in
section three. The combination of both effects due to MDR and the
generalized EUP (GEUP) is also presented.

\section{uncertainty principles and hawking temperature}
It has been argued in \cite{AH,Adler:2001vs} that there is an
intrinsic uncertainty about the Schwarzschild radius $R$ for those
photons in the vicinity of horizon. With the use of this fact one
can heuristically derive the the Hawking temperature of
Schwarzschild black holes. However, as pointed out in Ref.
\cite{Amati:1988tn,Konishi:1989wk}, the usual uncertainty principle
can not be naively applied to the space with large length scales
like as in (A)dS space. As a result, to properly obtain the
temperature of (A)dS Schwarzschild black holes in this manner the
usual uncertainty relations should be extended to include the
effects of the cosmological constant, which may be named as the
extended uncertainty principle (EUP) \cite{Bolen:2004sq,
Park:2007az}. It can be written as
\begin{equation}
\Delta x\Delta p\geq 1+\beta ^2 \frac{(\Delta x)^2}{L^2}
,\label{eup}
\end{equation}
where $\beta$ is a dimensionless real constant of order one, and $L$
is the characteristic large length scale related to the cosmological
constant as $\Lambda\sim 1/L^2$. In contrast to the ordinary
uncertainty relation, in EUP there exists an absolute minimum for
the momentum uncertainty
\begin{equation}
\Delta p\geq \frac{1}{\Delta x}+\frac{\beta ^2 \Delta x}{L^2}\geq
\frac{2\beta}{L}. \label{mu}
\end{equation} However, it is easy to see
that for a very large $L$, Eq.(\ref{eup}) goes back to the usual
Heisenberg¡¯s uncertainty relation.

Now, we consider a $4$-dimensional AdS black hole with the metric
\begin{equation}
ds^2=-N^2dt^2+N^{-2}dr^2+r^2(d\theta^2+\sin^2\theta d\phi^2),
\end{equation}
where
\begin{equation}
N^2=1+\frac{r^2}{L^2}-\frac{2GM}{r}.
\end{equation}

The event horizon $r_+$ can be obtained by setting $N^2=0$. Now,
using the standard results in statistical mechanics we expect that
the energy of photons emitted from the horizon can be identified
as the characteristic temperature of the Hawking radiation,
namely\cite{Adler:2001vs, Chen:2002tu}
\begin{equation}
T\sim E=p,\label{te}
\end{equation}
where $p$ is the momentum of photons emitted from the horizon.
Next, we propose that photons emitted from the black hole satisfy
the extended uncertainty principle(EUP). By modelling a black hole
as a black box with linear size $r_+$, the position uncertainty
$\Delta x$ of photons emitted from the black hole is just the
horizon $r_+$, i. e.
\begin{equation}
\Delta x\sim r_+.\label{li}
\end{equation}
 while the momentum of photons in a quantum mechanical region
 approximately satisfy the relation $p\sim \Delta p$. Then with (\ref{mu}),(\ref{te})and (\ref{li}), we immediately
obtain the Hawking temperature
\begin{equation}
T_{AdS}=\frac{1}{4\pi}
\left[\frac{1}{r_+}+\frac{3r_+}{L^2}\right],\label{ads}
\end{equation}
where a ``calibration factor'' $4\pi$ is introduced
\cite{Adler:2001vs} and the parameter $\beta^2$ is set to $3$ for
four dimensional black holes .

It is straightforward to obtain the temperature for dS black holes
by setting $L^2\rightarrow -L^2$ since the cosmological constant
$\Lambda \sim 1/L^2$ \cite{hha},
\begin{equation}
T_{dS}=\frac{1}{4\pi}
\left[\frac{1}{r_+}-\frac{3r_+}{L^2}\right].\label{ds}
\end{equation}

As pointed out in \cite{Park:2007az}, it is easy to see that in the
AdS case, the temperature has an absolute minimum due to EUP, while
in dS case, there exists a maximal radius for black hole horizon but
no minimal one. Nevertheless, in both cases the temperature will
suffer from the divergency as the size of the horizon approaches to
zero. This is a unsatisfactory point implying that the conventional
picture of Hawking radiation may not be applicable to the late stage
of black hole evaporation. To provide a more reasonable picture for
this, in next section we propose to modify the usual dispersion
relation for photons and discuss its possible impact on the
thermodynamics of (A)dS black holes.

\section{the impact of MDR on black hole physics}

It has been studied that the existence of a minimum length can
prevent black holes from total evaporation
\cite{Adler:2001vs,Cavaglia:2004jw}, where the generalized
uncertainty principle(GUP) plays an essential role. In this
section, we show that MDR may provide a similar mechanism to
describe the late stage of (A)dS black hole radiation in a
reasonable manner. As far as the uncertainty relations is
concerned, we will first consider the EUP case and then turn to
the GEUP one.
\subsection{EUP case}
After setting $m_0=0$ in (\ref{mdr1}) for photons we may rewrite
the general form of modified dispersion relations as
\begin{equation}
E=\frac{f_2(E;\eta)}{f_1(E;\eta)}p.
\end{equation}
As discussed in previous section, we expect that the energy of
photons emitted from black holes can be identified as the Hawking
temperature, but with MDR this identification will lead to a
recursion relation for the black hole temperature
\begin{equation}
T\sim
E=\frac{f_2(E;\eta)}{f_1(E;\eta)}p=\frac{f_2(T;\eta)}{f_1(T;\eta)}p.\label{mdr1'}
\end{equation}

Similarly, through the EUP in Eq.(\ref{mu}) the Hawking temperature
of (A)dS black holes can be written as
\begin{equation}
T=\frac{f_2(T;\eta)}{f_1(T;\eta)}T_0,\label{mdrt}
\end{equation}
where $T_0$ is given by \begin{equation} T_0=\frac{1}{4\pi}
\left[\frac{1}{r_+}\pm \frac{3r_+}{L^2}\right].\label{ori}
\end{equation} It is easy to see that for the
low energy case both functions $f_1$ and $f_2$ approach to one, the
temperature in Eq.(\ref{mdrt}) goes back to the original one.
However, at high energy level the temperature will receive important
corrections and the expression depends on the specific form of $f_1$
and $f_2$. For explicitness, we take the ansatz $f_1^2=1-(l_pE)^2$
and $f_2^2=1$. Then Eq.(\ref{mdrt}) becomes
\begin{equation}
T^2=\frac{1}{1-(l_pE)^2}T_0^2=\frac{1}{1-(l_pT)^2}T_0^2.
\end{equation}
From this equation we obtain the Hawking temperature as
\begin{equation}
T=\left[\frac{M_p^2}{2}\left(1-\sqrt{1-\frac{4T_0^2}{M_p^2}}\right)\right]^{1/2},\label{teup}
\end{equation}
where $M_p=l_p^{-1}$ (The other solution with plus sign ahead of the
square root is ruled out as it does not provide reasonable physical
meanings). From this equation it is easy to see that for large black
holes where $T_0\ll M_p$, the modified temperature goes back to the
usual one $T\sim T_0$. However, when the temperature increases with
the evaporation, we find that the modified temperature has a upper
limit $T\leq M_p/\sqrt 2$ and the inequality saturates when $T_0=
M_p/2$. The corresponding radius of the black hole horizon reaches a
minimal value $r_+=l_p/2\pi$. This situation is similar to the case
presented in \cite{Ling:2005bq} while the difference here is that
due to EUP, $T\sim T_0$ is also bounded from below with a minimum
value for large black holes. The existence of the minimal size of
horizon implies that black hole maybe have a final stable state and
this can be testified by calculating the heat capacity of the (A)dS
black holes, which turns out to be
\begin{equation}
C_{(A)dS}=-\frac{2\pi\sqrt{1-4T^2_0/M^2_p}\left(1\pm\frac{3r^2_+}{L^2}\right)}
{G\sqrt{1-T^2/M^2_p}\left(\frac{1}{r^2_+}\mp\frac{3}{L^2}\right)},\label{eq29}
\end{equation}
where the upper sign and lower sign correspond to AdS and dS black
holes respectively. Obviously the heat capacity vanishes as $T_0=
M_p/2$.

As a result, we argue that MDR may provide a mechanism to prevent
(A)dS black holes from total evaporation such that an explosive
disaster of black holes can be avoided.

In the end of this section we briefly discuss the correction to the
entropy of (A)dS black holes due to MDR. It is expected that the
first thermodynamical law still holds for large black holes (i.e.
$8G\ll A$). Plugging Eq.(\ref{teup}) into $dM=TdS$, we have
\begin{equation}
dS=\frac{1}{2\sqrt{G}}\left[A-A\sqrt{1-\frac{8G}{A}\left(1+\frac{3A}{4\pi
L^2}\right)^2}\right]^{-\frac{1}{2}}\left(1+\frac{3A}{4\pi
L^2}\right)dA,\label{entropy}
\end{equation}
where $G=1/(8\pi M_p^2)$ and $A=4\pi r_+^2$ is the area of the
horizon. Eq.(\ref{entropy}) can be written in a simpler way,
\begin{equation}
dS=\frac{1}{4\sqrt{2}G}\left[1+\sqrt{1-\frac{8G}{A}\left(1+\frac{3A}{4\pi
L^2}\right)^2}\right]^{1/2}dA.
\end{equation}
Now using the condition for large black holes $8G\ll A\ll L^2$, we
can obtain the entropy of the black hole,
\begin{equation}
S\simeq\frac{A}{4G}-\frac{1}{4}\ln \frac{A}{4G}-\frac{3A}{8\pi
L^2}-\frac{9A^2}{128\pi^2L^4}.
\end{equation}
The first term is the conventional Bekenstein-Hawking entropy while
the other terms are corrections due to MDR. It is interesting to
notice that the entropy has a logarithmic term, agreement with the
results obtained in string theory and loop quantum gravity
\cite{Strominger:1996sh,Solodukhin:1997yy,Rovelli:1996dv,Ashtekar:1997yu,Kaul:2000kf}.

\subsection{GEUP case}
In previous section we have demonstrated that an appropriate form
of MDR will provide a cutoff for the temperature of (A)dS black
holes. Definitely we may consider this effect in the context of
generalized EUP(GEUP), which is given by \cite{Park:2007az}
\begin{equation}
\Delta x\Delta p\geq 1+\alpha ^2l_p^2(\Delta p)^2\pm\beta ^2
\frac{(\Delta x)^2}{L^2},
\end{equation}
where $\alpha$ is a new dimensionless parameter with a $``\pm''$
sign corresponding to AdS and dS black holes respectively. As
shown in \cite{Park:2007az}, for GEUP one has the inequality
\begin{equation}
\Delta p^{(-)}\leq\Delta p\leq\Delta p^{(+)},
\end{equation}
where
\begin{equation}
\Delta p^{(\pm)}=\frac{\Delta x}{2\alpha ^2l_p^2}\left[1\pm
\sqrt{1-\frac{4\alpha ^2l_p^2}{(\Delta x)^2}\left[1\pm\beta
^2\frac{(\Delta x)^2}{L^2}\right]}\right].\label{18}
\end{equation}
Then, one can find that $\Delta x$ has the absolute minimum
\begin{equation}
(\Delta x)^2\geq\frac{4\alpha ^2l_p^2}{1\mp 4\alpha ^2\beta
^2l_p^2/L^2}.\label{19}
\end{equation}

Now closely following the procedure in previous section, one can
obtain a modified Hawking temperature as
\begin{equation}
T=\frac{f_2}{f_1}T_{GEUP},\label{tgeup}
\end{equation}
where
\begin{equation}
T_{GEUP}=\frac{1}{4\pi}\frac{r_+}{2\alpha
^2l_p^2}\left[1-\sqrt{1-\frac{4\alpha
^2l_p^2}{r_+^2}\left[1\pm\frac{3r_+^2}{L^2}\right]}\right],
\label{21}\end{equation} with $\beta^2=3$. We notice that
Eq.(\ref{tgeup}) is just an extension of Eq.(\ref{mdrt}) with
$T_0\rightarrow T_{GEUP}$. In hence, if one fixes both functions
$f_1$ and $f_2$ as given in the previous section, the final
modified temperature of black holes becomes
\begin{equation}
T_{(A)dS}=\left[\frac{M_p^2}{2}\left(1-\sqrt{1-\frac{4T_{GEUP}^2}{M_p^2}}\right)\right]^{1/2}.\label{22}
\end{equation}

Comparing Eqs.(\ref{21}) and (\ref{22}), it is interesting to
notice that this specific form of $MDR$ leads to a similar
modification to the temperature as $GEUP$. Thus, both $MDR$ and
$GEUP$ may independently provide a upper limit for the temperature
of black holes. In the context of $GEUP$ this maximal value is
given by
\begin{equation}
T^{max}_{GEUP}=\frac{1}{4\pi\alpha l_p}\frac{1}{(1\mp
12\alpha^2l_p^2/L^2)^{1/2}},
\end{equation}
where $\alpha$ is supposed to be non-zero and order one
($\alpha\rightarrow 0$ leading to $T_{GEUP}\rightarrow T_0$ as seen
from Eq.(\ref{21})). Our above discussion indicates that the
combination of $MDR$ and $GEUP$ does not change the whole picture
obtained in the subsection of $EUP$ case greatly, but provides
further modifications to the final temperature as well as the mass
of black holes.

\section{summary and discussions}
In this paper we have investigated the impact of modified
dispersion relations on the thermodynamics of (A)dS black holes.
We have shown that MDR contributes corrections to the usual
Hawking temperature as well as the entropy of black holes. Such
corrections may play an important role in the understanding of the
final fate of black holes. In particular, it provides a vanishing
heat capacity at the late stage of black hole evaporation such
that it can prevent black holes from total evaporation, but
arriving a stable state with a minimum horizon radius and the
remnant can be treated as a candidate for cold dark matter.

It is worthwhile to point out that through the paper we identify
 the expectation value of the energy as the temperature of black holes,
which is a standard relation in statistics for radiation
particles. However, as pointed out in \cite{Alexander:2001ck} and
\cite{Ling:2006az}, MDR may also change the statistical properties
of the ensemble such that $E\sim T$ is not strictly satisfied. It
may be modified as $E\sim T(1+\delta lp^2T^2)$. Therefore, more
exact results of the Hawking temperature maybe have to take these
modifications into account. However, a delicate calculation shows
that this modification will not change the main picture about the
final fate of black holes as we present here.

Through the paper, we choose a specific form of MDR in which both
energy and momentum of particles are bounded. It is completely
possible to extend our strategy to other $MDR$ forms and a parallel
analysis should be straightforward if such forms could provide a
upper limit for the energy of particles as expected from the side of
doubly special relativity. Furthermore, our discussions are
applicable to (A)dS black holes in higher dimensional spacetime once
the parameter $\beta$ in $EUP$ is properly fixed.

\begin{acknowledgments}
Xin Han would like to thank Wei Tang for helpful discussions.  We
are also grateful to all members at the Center for Relativistic
Astrophysics and High Energy Physics of Nanchang University for
generous support which made this work possible. This work is partly
supported by NSFC(Nos. 10405027,  10663001), JiangXi SF (Nos. 0612036, 0612038), the key project of Chinese Ministry of Education (No. 208072) and Fok Ying Tung Eduaction Foundation(No.
111008).  We also acknowledge the support by the Program for
Innovative Research Team of Nanchang University.
\end{acknowledgments}

\bibliography{apssamp}

\end{document}